\documentclass[epj]{svjour}
\usepackage{xcolor}
\usepackage{graphicx}
\usepackage{amsmath}
\begin{document}
\title{Systematic Comparison of Tsallis Statistics for Charged Pions Produced In $pp$ Collisions}
\author{A.S.~Parvan\inst{1,2}
\and O.V.~Teryaev\inst{1}
\and J.~Cleymans\inst{3}
}                     
%
%
\institute{Bogoliubov Laboratory of Theoretical Physics, Joint Institute for Nuclear Research, Dubna,
Russian Federation \and
Department of Theoretical Physics, Horia Hulubei National Institute of Physics and Nuclear Engineering,
Bucharest-Magurele, Romania
\and
UCT-CERN Research Centre and Department of Physics, University of Cape Town, South Africa}
\date{Received: date / Revised version: date}
%
\abstract{
The energy dependence of Tsallis statistics parameters is presented for charged pions produced at
beam energies ranging between 6.3 GeV
and 7 TeV. It is
found that deviations from Boltzmann statistics are monotonically
growing with beam energy.
The energy dependence of the parameters $T$ and $q$  of the negatively charged pions in the 
energy range $6.3<\sqrt{s}<7000$ GeV reveals that
the deviation of the transverse momentum distribution from the exponential 
becomes more and more pronounced as the beam energy is increased. 
The parameter $q$ increases with
beam energy while  the temperature $T$ slowly decreases.
\PACS{
      {13.85.-t}{ Hadron-induced high- and super-high-energy interactions}   \and
      {13.85.Hd}{Inelastic scattering: many-particle final states} \and
      {24.60.-k}{Statistical theory and fluctuations}
     } 
} 
\titlerunning{Systematic Comparison of Tsallis Statistics}
\authorrunning{A.S.~Parvan, O.V.~Teryaev, J.~Cleymans}
\maketitle

\section{Introduction}
The use of non-extensive distributions (commonly referred to as Tsallis distributions~\cite{tsallis}) 
in high-energy physics  has a long history \cite{Bediaga00,Beck00,Wilk00,Walton00,Alberico00,Zimanyi05,Trainor08,Wilk09,Biro09}.
They have been surprisingly successful
in describing transverse momentum distributions in p-p collisions at the LHC
At first these were used as a parametrization of the experimental 
data~\cite{STAR1,STAR2,PHENIX1,PHENIX2,ALICE1,ALICE2,ALICE3,CMS1,CMS2,CMS3,ATLAS}
and, more recently, they have  attracted considerable attention in theoretical 
papers~\cite{biro,zheng1,gao,zheng2,zheng3,wilk,marques,urmossy,sorin,ryb}.
The parametrization holds extremely well even in the 
high transverse momentum region traditionally reserved for perturbative QCD as shown in~\cite{wong1,wong2,wilk2,azmi,azmi-cleymans}.
This is indeed unexpected and should be viewed with some skepticism, it definitely calls for further investigations.

In the present paper we 
 extend the analysis to  lower beam energies and include results
from the  NA61 collaboration~\cite{NA61} for negatively charged pions measured at $p_{\text{lab}}$ = 20, 31, 40. 80 and 158 GeV/c,
as well as results from the STAR~\cite{STAR1,STAR2},  PHENIX~\cite{PHENIX1,PHENIX2}, ALICE~\cite{ALICE1,ALICE2}, 
ATLAS~\cite{ATLAS} and CMS~\cite{CMS1,CMS2,CMS3} collaborations.

The energy dependence of the parameters $T$ and $q$  of the negatively charged pions in the 
energy range $6.3<\sqrt{s}<7000$ GeV reveals that
the deviation of the transverse momentum distribution from the exponential 
becomes more and more pronounced as the beam energy is increased. 
The parameter $q$ increases with
beam energy while  the temperature $T$ slowly decreases.
\section{Transverse momentum distributions in $pp$ collisions}
The experimental results on the transverse momentum distributions for negatively charged pions
are fitted to the Tsallis distribution as  used in
Ref.~\cite{sorin}. 
The explicit form used is given by the following expression
\begin{eqnarray}\label{1}
\frac{d^{2}N}{dp_{T}dy} &=& gV\frac{p_{T}m_{T}\cosh y}{(2\pi)^2} \nonumber \\
 && \times \left[1+(q-1)\frac{m_{T}\cosh y -\mu}{T}\right]^{q/(1-q)},
\end{eqnarray}
where $m_{T}=\sqrt{p_{T}^2+m^{2}}$, $m$ is the particle rest mass, $\mu, T, V$  are the chemical potential, temperature,
and  volume, $g$ is the spin degeneracy factor and $q$ is a parameter whose range will be discussed in detail below.
In the limit where the parameter $q\to 1$
one recovers the
Maxwell-Boltzmann distribution:
\begin{equation}\label{2}
\frac{d^{2}N}{dp_{T}dy} = gV\frac{p_{T}m_{T}\cosh y}{(2\pi)^2} \ e^{-\frac{m_{T}\cosh y -\mu}{T}}.
\end{equation}

Obviously the integral of Eq.~(\ref{1}) over all transverse momenta and rapidities has to exist in order to produce
a finite number of particles, this leads to the constraint that  $q < 3/2$.
For massless particles the integral can be calculated analytically, the  zero chemical potential case is given by:
\begin{equation}
n =\ \frac{g\ T^3}{\pi^2}\ \ \frac{1}{(2 - q)\ (3 - 2q)}.
\end{equation}
While the energy density is given by:

\begin{equation}
\epsilon =\ \frac{g\ 3T^4}{\pi^2}\ \ \frac{1}{(2-q)\ (3 - 2q)\ (4 - 3q)}.
\end{equation}
which leads to a more stringent condition
\begin{equation}
q < \frac{4}{3}   .
\end{equation}

In contrast to the exact Tsallis statistics~\cite{tsallis}, the single-particle distribution function~(\ref{1}) of the
Tsallis-factorized statistics is represented in a simple explicit form and serves as a powerful tool to study the
experimental data on the transverse momentum distributions. The single-particle distribution function of the
Tsallis statistics cannot be written in an explicit form~(\ref{1}) because the many-body distribution function of the
exact Tsallis statistics does not factorize into the product of the single-particle distribution
functions~\cite{Parvan15}. Only in the factorization (dilute gas) approximation the single-particle distribution
functions of the Tsallis statistics can be written explicitly~\cite{Buyukkilic93}.  For simplicity we will refer
to Eq.~\ref{1} as Tsallis distribution.

For a given rapidity range $y_{0}<y<y_{1}$ the transverse momentum distribution can be written as
\begin{eqnarray}\label{3}
\left.\frac{d^{2}N}{dp_{T}dy}\right|_{y_0<y<y_1} &=& gV\frac{p_{T}m_{T}}{(2\pi)^2} \int\limits_{y_{0}}^{y_{1}} dy  \cosh y \nonumber \\
 \times && \left[1+(q-1)\frac{m_{T}\cosh y -\mu}{T}\right]^{q/(1-q)}. \;\;\;\;
\end{eqnarray}
At mid-rapidity $y=0$ and chemical potential $\mu=0$ the distribution function~(\ref{1})
reduces to
\begin{equation}\label{4}
\frac{d^{2}N}{dp_{T}dy} = gV\frac{p_{T}m_{T}}{(2\pi)^2} \ \left[1+(q-1)\frac{m_{T}}{T}\right]^{q/(1-q)}.
\end{equation}

\begin{figure}
\includegraphics[width=0.49\textwidth]{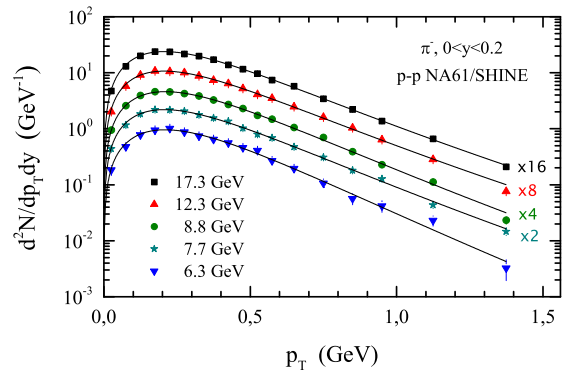} \vspace{-0.3cm}
\caption{(Color online) Transverse momentum distribution of negatively charged pions $\pi^{-}$ produced
in $pp$ collisions as obtained by the NA61/SHINE Collaboration~\cite{NA61} at $\sqrt{s}=6.3,7.7,8.8,12.3$ and $17.3$ GeV
in the rapidity interval $0<y<0.2$. The solid curves are the fits of the data to the
Tsallis distribution (\ref{3}).} \label{fig1}
\end{figure}
Let us study the changes in the transverse momentum distribution of the negatively charged pions $\pi^{-}$ (selected just because the data
are available at all the considered energies) are  produced
in $pp$ collisions with energy in the range from $\sqrt{s}=6.3$ GeV to $7000$ GeV. Figure~\ref{fig1} represents the
transverse momentum distribution of $\pi^{-}$ pions produced in the $pp$ collisions as obtained by the NA61/SHINE
Collaboration~\cite{NA61} at $\sqrt{s}=6.3,7.7,8.8,12.3$ and $17.3$ GeV
(which are close
to the energy range of the future NICA collider)
in the rapidity interval $0<y<0.2$. The symbols represent the experimental data. The solid curves are the
fits
to the experimental data
using
 the Tsallis function (\ref{3}). The values of the parameters 
are given in Table~\ref{t1}. The Tsallis function describes very well
the experimental data. At the energies of the NA61/SHINE Collaboration the transverse momentum distribution of
negatively charged pions $\pi^{-}$ has the power law form, however, its deviation from the exponential function is not
so large. The values of the parameter $q$ are close to unity. Herewith, the temperature $T$
is
approximately $T\sim 100$ MeV. Moreover, the value of the radius $R$ is large in comparison to the geometrical sizes of
the system composed from two protons. Note that the experimental data of the NA61/SHINE Collaboration were also fitted
to the Tsallis function in Ref.~\cite{ryb},
we can reproduce these results within the limits of errors at $y=0$.

\begin{figure}
\includegraphics[width=0.48\textwidth]{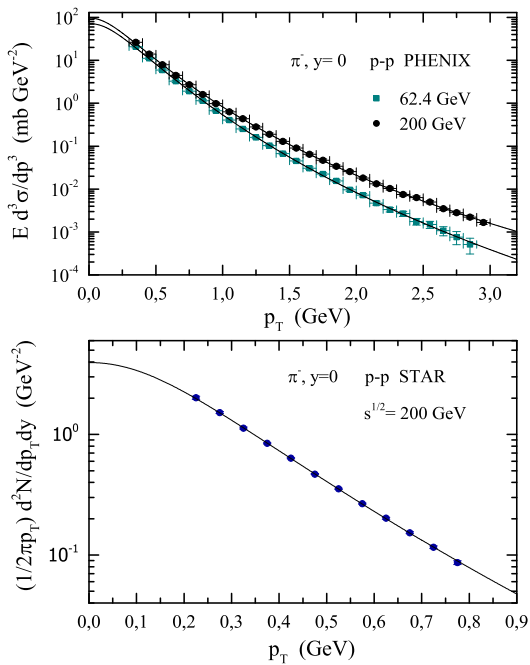} 
\caption{
(Color online) Transverse momentum distribution of negatively charged pions $\pi^{-}$ produced in 
$pp$ collisions as obtained by the PHENIX Collaboration~\cite{PHENIX1,PHENIX2} at $\sqrt{s}=200$ and $62.4$ GeV at 
mid-rapidity (upper panel) and by the STAR Collaboration~\cite{STAR1,STAR2} at $\sqrt{s}=200$ GeV in the minimum 
bias (lower panel). The solid curves are the fits of the data to the Tsallis distribution (\ref{5}) at 
rapidity $y=0$.
} 
\label{fig2}
\end{figure}

\begin{figure}
\includegraphics[width=0.48\textwidth]{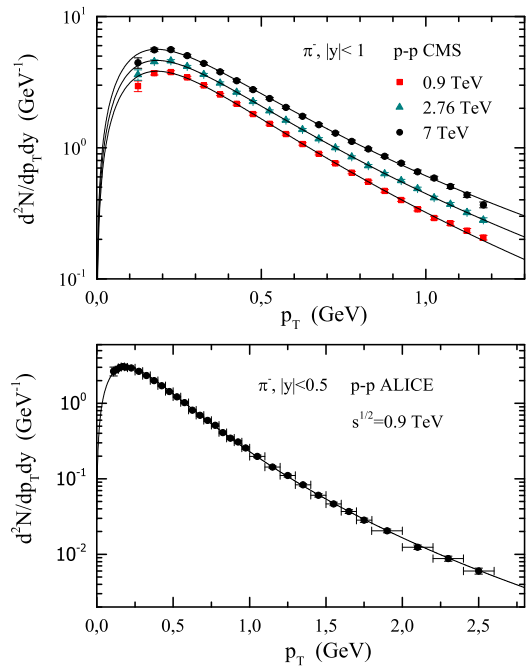} 
\caption{(Color online) Transverse momentum distribution of negatively charged pions $\pi^{-}$ produced in $pp$ collisions
as obtained by the CMS Collaboration~\cite{CMS1,CMS2,CMS3} at $\sqrt{s}=0.9,2.76$ and $7$ TeV in the rapidity
interval $|y|<1$ (upper panel) and by the ALICE Collaboration~\cite{ALICE2} at $\sqrt{s}=0.9$ TeV in the
rapidity interval $|y|<0.5$ (lower panel). The solid curves are the fits of the data to the Tsallis
distribution (\ref{3}).} \label{fig3}
\end{figure}

Figure~\ref{fig2} presents the transverse momentum distributions of $\pi^{-}$ pions produced in the proton-proton collisions as obtained
by the PHENIX Collaboration~\cite{PHENIX1,PHENIX2} at $\sqrt{s}=200$ and $62.4$ GeV at mid-rapidity and by the STAR
Collaboration~\cite{STAR1,STAR2} at $\sqrt{s}=200$ GeV. The symbols represent the experimental data of the 
PHENIX and STAR Collaborations. The solid
curves are the fits of the experimental data to the 
Tsallis function (\ref{4}) divided by  the geometrical factor $2\pi p_T$:
\begin{equation}\label{5}
\frac{1}{2\pi p_{T}}\frac{d^{2}N}{dp_{T}dy} = gV\frac{m_{T}}{(2\pi)^3} \ \left[1+(q-1)\frac{m_{T}}{T}\right]^{q/(1-q)}.
\end{equation}
The values of the parameters of this Tsallis function are given in Table~\ref{t1}.
The experimental data for the transverse momentum distributions of the negatively charged pions $\pi^{-}$ are very well
described by the function (\ref{5}). The experimental transverse momentum distributions of $\pi^{-}$ pions measured by the
PHENIX and STAR Collaborations  differ from the exponential function. The values of the parameter $q$ are greater than unity.
\begin{figure}
\includegraphics[width=0.48\textwidth]{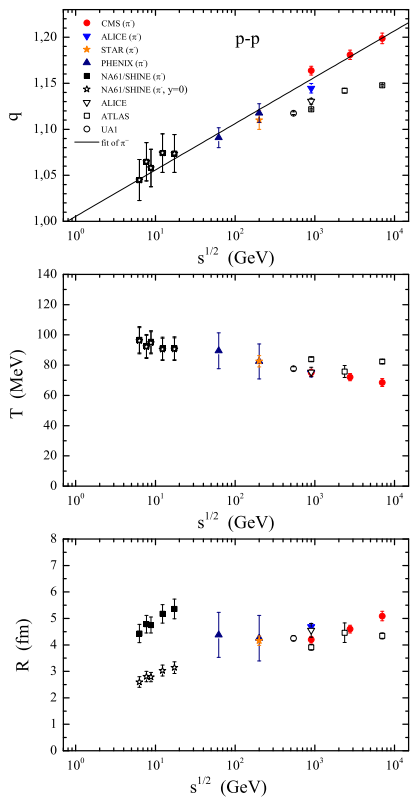} \vspace{0.1cm}
\caption{(Color online) Energy dependence of the temperature $T$, radius $R$ and the parameter $q$ of the Tsallis
distribution. Solid points are results of the fit for the negatively charged pions $\pi^{-}$ produced in $pp$ collisions
as obtained by the NA61/SHINE~\cite{NA61}, PHENIX~\cite{PHENIX1,PHENIX2}, STAR~\cite{STAR1,STAR2}, ALICE~\cite{ALICE2} and
CMS~\cite{CMS1,CMS2} Collaborations.  Open stars are results of the fit at $y=0$ for the data of NA61/SHINE.
Open squares, triangles and circles are results of the fit for the charged hadron yields produced
in $pp$ and $p\bar{p}$ collisions as obtained by the ATLAS, ALICE and UA1 Collaborations, respectively
(taken from Ref.~\cite{sorin}). Data for the parameter $q$ for $\pi^{-}$ were fitted
by Eq.~(\ref{6}).} \label{fig4}
\end{figure}

Figure~\ref{fig3} represents the transverse momentum distributions of the negatively charged pions $\pi^{-}$ produced in
 $pp$ collisions as obtained by the CMS Collaboration~\cite{CMS1,CMS2} at $\sqrt{s}=0.9,2.76$ and $7$ TeV in the rapidity
interval $|y|<1$ and by the ALICE Collaboration~\cite{ALICE2} at $\sqrt{s}=0.9$ TeV in the rapidity
interval $|y|<0.5$. The symbols represent the experimental data. The solid curves are the fits of the experimental
data to the Tsallis function (\ref{3}); the values of the parameters of the Tsallis function (\ref{3}) are given
in Tab.~\ref{t1}. The experimental data of the CMS Collaboration for the transverse momentum distributions of the
negatively charged pions are described very well by the power law function (\ref{3}) with exception only of the
lowest $p_{T}$ data (see the upper panel of
Fig.~\ref{fig3}). The experimental data of the ALICE Collaboration at $\sqrt{s}=0.9$ TeV for $\pi^{-}$ pions are
described very well by the function (\ref{3}) in all $p_{T}$ region (see the lower panel of Fig.~\ref{fig3}).
Thus the experimental transverse momentum distributions of $\pi^{-}$ pions measured by the ALICE and CMS
Collaborations cannot be described by the exponential function.

The energy dependence of the parameters of the Tsallis distribution for the negatively charged pions $\pi^{-}$ produced
in $pp$ collisions is shown in Fig.~\ref{fig4}. The parameters of $\pi^{-}$ pion distributions are compared with the
parameters of the charged hadron distributions given in Ref.~\cite{sorin}. Solid points are results of the fit for
the negatively charged pions $\pi^{-}$. Open points are results of the fit for the charged hadron distributions produced
in $pp$ and $p\bar{p}$ collisions as obtained by the ATLAS, ALICE and UA1 Collaborations, respectively~\cite{sorin}.
Open stars are results of the fit at $y=0$ for the data of NA61/SHINE Collaboration. The values of the parameters
are given in the Table~\ref{t1} and~\ref{t2}.

It is clearly seen that the transverse momentum distribution of $\pi^{-}$ pions has a power law form and increasingly
deviates from the exponential function with collision energy (the parameter $q$ is not equal to unity and increases
with $\sqrt{s}$).
Within the range of beam energies considered here the dependence of $q$ on $\sqrt{s}$ is linear on a logarithmic scale.
To show explicitly this dependence we parametrize the parameter $q$ for $\pi^{-}$ pions by
the function of the form
\begin{equation}\label{6}
  q=1 + \ln\left(\frac{\sqrt{s}}{\sqrt{s_{0}}}\right)^{a_{0}},
\end{equation}
where
 $a_{0}=0.0219\pm 0.0014$
and $\sqrt{s_{0}}=0.7805\pm 0.3673$ GeV. Eventually this rise will
come to a halt and reach the asymptotic limit 4/3 discussed earlier
in this paper. For the suggested linear parametrization this happens at
extremely high energies, being even higher than the highest
energies of  protons in  cosmic rays. \\
At NA61/SHINE energies the parameter $q$
for $\pi^{-}$ pions is close to unity.
Let us stress that the use of the pion data instead of data for charged hadrons first used in \cite{sorin} and
explored in \cite{ryb} allows us to obtain a much better fit than in previous analyses.

With increasing beam  energy, as attained in  PHENIX, STAR, ALICE and CMS, the parameter
$q$ for $\pi^{-}$ pions clearly increases and becomes significantly different from unity.
Note that the values of the parameter $q$ for $\pi^{-}$ pions measured by the CMS Collaboration differ from the
values of the parameter $q$ of charged hadrons obtained by the ALICE and ATLAS Collaborations.

As seen in the top panel of Fig.~\ref{fig4}, the temperature $T \sim 100$ MeV of
$\pi^{-}$ pions  slowly decreases with the energy of the collision. The
greater the deviation of the distribution function from the
exponential function, the smaller the temperature is. 
The values of the temperature $T$ for $\pi^{-}$
pions measured at PHENIX, STAR and ALICE energies are significantly
smaller than the values of the temperature $T$ of the negatively
charged pions obtained at NA61/SHINE energies. Note that the values
of the temperature $T$ of the charged hadrons measured by the ATLAS
Collaboration given in Ref.~\cite{sorin} differ from the values
of the temperature $T$ of $\pi^{-}$ pions of the CMS Collaboration
obtained by the Tsallis distribution (\ref{3}).
For comparison, using the standard blast-wave description, as used e.g. in~\cite{prl2012} leads to a similar value
kinetic freeze-out temperature of $T_{kin} = 95 \pm 10$ MeV for an average transverse expansion velocity of $<\beta_T> = $ 0.65$\pm$ 0.02
albeit for heavy-ion collisions. 

One can also attribute such energy behaviour to the interplay of longitudinal and transverse degrees of freedom. The temperature in the longitudinal $p_{L}$ space is of the order of tens of GeV and depends on the available energy and multiplicity of secondaries~\cite{Wilk09}. 
Thus, a slight decrease in temperature with energy may be explained by the fact that most part of the energy of the system flows in the longitudinal direction. This may be considered as a sort of implementation of suggestion ~\cite{Tsallis02}, that "higher collision energies do not increase the transverse momenta temperature but increase instead the number of involved bosons that are produced (like water boiling at higher flux of energy, where only the rate of vapor production is increased, but not the temperature)".

As seen in the middle panel of Fig.~\ref{fig4}, the radius $R$ for $\pi^{-}$ pions is
only mildly dependent on
the energy of
the collision. The values of the radius $R$ of the system for $\pi^{-}$ pions differ essentially from the geometrical
sizes of the compound system of two protons. However, the values of the radius $R$ for $\pi^{-}$ pions are consistent
with the values of the radius $R$ for the charged hadrons given in Ref.~\cite{sorin}. It is interesting that maximal value of radius appears in the same energy region as famous "horn" in the kaons relative multiplicity.

\begin{table*}
\begin{tabular}{cccccc}
 \hline
 \hline
 $\quad$ Collaboration $\quad$ & $\quad\sqrt{s}$, GeV $\quad$ & $\qquad$ $T$, MeV$ \qquad$ & $\qquad$ $\qquad$ $R$, fm$\qquad$ & $\qquad$ $\qquad$ $q$ $\qquad$ & $\qquad$  $\chi^{2}/ndf$ $\qquad$  \\
 \hline
 NA61/SHINE     & 6.3       & 96.76$\pm$8.69  & 4.431$\pm$0.344   & 1.0449$\pm$0.0223 & 2.704/15  \\
 NA61/SHINE     & 7.7       & 92.68$\pm$7.67  & 4.782$\pm$0.334   & 1.0647$\pm$0.0208 & 1.140/15  \\
 NA61/SHINE     & 8.8       & 95.39$\pm$7.33  & 4.749$\pm$0.301   & 1.0580$\pm$0.0204 & 0.989/15  \\
 NA61/SHINE     & 12.3      & 91.03$\pm$7.43  & 5.172$\pm$0.350   & 1.0741$\pm$0.0209 & 0.891/15  \\
 NA61/SHINE     & 17.3      & 91.17$\pm$7.56  & 5.358$\pm$0.375   & 1.0736$\pm$0.0205 & 0.459/15  \\
 PHENIX         & 62.4      & 89.52$\pm$11.83  & 4.379$\pm$0.853   & 1.0909$\pm$0.0108 & 0.938/23  \\
 PHENIX         & 200.0     & 82.50$\pm$11.56  & 4.255$\pm$0.861   & 1.1177$\pm$0.0101 & 0.758/24  \\
 STAR           & 200.0     & 82.57$\pm$3.74  & 4.159$\pm$0.173   & 1.1100$\pm$0.0100 & 2.738/9  \\

 ALICE          & 900.0     & 74.69$\pm$2.37  & 4.686$\pm$0.125   & 1.1446$\pm$0.0051 & 2.183/30  \\

 CMS            & 900.0     & 75.02$\pm$2.30  & 4.198$\pm$0.124   & 1.1637$\pm$0.0048 & 11.080/19  \\
 CMS            & 2760.0    & 72.10$\pm$2.38  & 4.597$\pm$0.144   & 1.1809$\pm$0.0051 & 7.425/19  \\
 CMS            & 7000.0    & 68.53$\pm$2.62  & 5.090$\pm$0.182   & 1.1987$\pm$0.0057 & 11.500/19  \\
\hline
\hline
\end{tabular}
\caption{Parameters of the Tsallis fit for $\pi^{-}$ mesons produced in $pp$ collisions at different energies.}
\label{t1}
\end{table*}

\begin{table*}
\begin{tabular}{ccccc}
 \hline
 \hline
 $\qquad\sqrt{s}$, GeV $\qquad$  & $\qquad$ $T$, MeV$ \qquad$ & $\qquad$ $\qquad$ $R$, fm$\qquad$ &$\qquad$ $\qquad$ $q$ $\qquad$ &  $\qquad$  $\chi^{2}/ndf$ $\qquad$  \\
 \hline
 6.3       & 96.12$\pm$8.64  & 2.597$\pm$0.201   & 1.0449$\pm$0.0223 & 2.704/15  \\
 7.7       & 92.05$\pm$7.62  & 2.803$\pm$0.196   & 1.0648$\pm$0.0208 & 1.140/15  \\
 8.8       & 94.76$\pm$7.28  & 2.784$\pm$0.177   & 1.0580$\pm$0.0204 & 0.989/15  \\
 12.3      & 90.44$\pm$7.39  & 3.031$\pm$0.205   & 1.0741$\pm$0.0209 & 0.892/15  \\
 17.3      & 90.57$\pm$7.52  & 3.140$\pm$0.220   & 1.0737$\pm$0.0205 & 0.459/15  \\
\hline
\hline
\end{tabular}
\caption{Parameters of the Tsallis fit at $y=0$ for $\pi^{-}$ pions measured at NA61/SHINE energies.}
\label{t2}
\end{table*}

\section{Discussion and conclusions}~\label{sec3}
The experimental data on the transverse momentum distribution of the negatively charged pions produced in $pp$ collisions at different energies were fitted to the Maxwell-Boltzmann distribution of the Tsallis statistics at
zero chemical potential. 
We have found the energy dependence of the parameters $T$, $R$ and $q$ of the Tsallis
distribution of the negatively charged pions in the energy range $6.6<\sqrt{s}<7000$ GeV. We have revealed that
the deviation of the transverse momentum distribution of negatively charged pions from the exponential function toward
the power law distribution increases with energy of $pp$ collisions. We have found that the parameter $q$ increases with
beam energy while  the temperature $T$ slowly decreases.
The radius $R$ in $pp$ collisions is a constant independent on the energy.
At small values of energy of collision, where the transverse momentum distribution of negatively charged pions
is close to the exponential function, the temperature is largest. It is interesting, that
parameter $q$ turns to $1$ at low (unphysical) energy surprisingly close to $\rho-$meson mass, which is the natural
hadronic scale.

%


{\bf Acknowledgments:}
We are indebted to  G.I. Lykasov and A.S. Sorin for stimulating discussions.
This work was supported in part by the joint research project and grant of JINR and IFIN-HH (protocol N~4543), by
the Program of South Africa-JINR collaboration, and by RFBR (grant 14-01-00647).

%

\end{document}